\documentclass{raa}
\usepackage{graphicx}
\usepackage{times}
\usepackage{amsfonts}
\usepackage{epsfig,natbib}
\usepackage{color}
\usepackage{amssymb,latexsym}

\def\br{{\mathbf{r}}}

\def\cmax{c_{max}}
\def\npmax{n_{p_{max}}}

\begin{document}

  \title{A Modified TreePM Code}
  
  \volnopage{Vol.0 (200x) No.0, 000--000}
  \setcounter{page}{1}          
  
  \author{Nishikanta Khandai
    \and J. S. Bagla
  }
  \institute{Harish-Chandra Research Institute,  Chhatnag Road, Jhusi,
    Allahabad 211019, India. \\
    e-mail: nishi@hri.res.in, jasjeet@hri.res.in
  }
  \date{Received~~2009 month day; accepted~~2009~~month day}

  \abstract{We discuss the performance characteristics of using 
    the modification of the tree code suggested by Barnes 
    \citep{1990JCoPh..87..161B} in the context of the TreePM code. 
    The optimisation involves identifying groups of particles and 
    using only one tree walk to compute force for all the 
    particles in the group.  
    This modification has been in use in our implementation of the TreePM
    code for some time, and has also been used by others in codes that 
    make use of tree structures.
    In this paper, we present the first detailed study of the performance
    characteristics of this optimisation.  
    We show that the modification, if tuned properly can speed up the 
    TreePM code by a significant amount.
    We also combine this modification with the use of individual time steps and
    indicate how to combine these two schemes in an optimal fashion. 
    We find that the combination is at least a factor of two faster than the
    modified TreePM without individual time steps. 
    Overall performance is often faster by a larger factor, as the
    scheme of groups optimises use of cache for large simulations.
    \keywords{Cosmology; Gravitation; Methods: numerical; 
    Large-scale structure of universe}
  }
  \authorrunning{Khandai \& Bagla}
  \titlerunning{Modified TreePM}

\maketitle

\section{Introduction}

Large scale structures traced by galaxies are believed to have formed by
amplification of small perturbations
\citep{1980lssu.book.....P,1999coph.book.....P,2002tagc.book.....P,2002PhR...367....1B}.   
Galaxies are highly over-dense systems, matter density $\rho$ in galaxies is
thousands of times larger than the average density $\bar\rho$ in the
universe.  
Typical density contrast ($\delta \equiv \rho/{\bar\rho} - 1$) in matter at
these scales in the early universe was much smaller than unity.  
Thus the problem of galaxy formation and the large scale distribution of
galaxies requires an understanding of the evolution of density perturbations
from small initial values to the large values we encounter today.

Initial density perturbations were present at all scales that have been
observed
\citep{2007ApJS..170..377S,2007ApJ...657..645P}.
The equations that describe the evolution of density perturbations in an
expanding universe have been known for several decades
\citep{1974A&A....32..391P} and these are easy to solve when the amplitude of
perturbations is small.   
Once density contrast at relevant scales becomes comparable to unity, 
perturbations becomes non-linear and coupling with perturbations at other
scales cannot be ignored.  
The equation for evolution of density perturbations cannot be solved for
generic initial conditions in this regime.
N-Body simulations (e.g., see
\citep{1985ApJS...57..241E,1998ARA&A..36..599B,1997Prama..49..161B,2005CSci...88.1088B})
are often used to study the evolution in this regime.  
Alternative approaches can be used if one requires only a limited
amount of information and in such a case either quasi-linear approximation
schemes 
\citep{2002PhR...367....1B},
\citep{1970A&A.....5...84Z,1989MNRAS.236..385G,1992MNRAS.259..437M,1993ApJ...418..570B,1994MNRAS.266..227B,1995PhR...262....1S,1996ApJ...471....1H}
or scaling relations
\citep{1977ApJS...34..425D,1991ApJ...374L...1H,1995MNRAS.276L..25J,2000ApJ...531...17K,1998ApJ...508L...5M,1994MNRAS.271..976N,1996ApJ...466..604P,1994MNRAS.267.1020P,1996MNRAS.278L..29P,1996MNRAS.280L..19P,2003MNRAS.341.1311S}
suffice. However, even the approximation schemes and scaling relations must be compared
with simulations before these can be used with confidence.

Last three decades have seen a rapid development of techniques and computing
power for cosmological simulations and the results of these simulations have
provided valuable insight into the study of structure formation.
The state of the art simulations used less than $10^5$ particles two
decades ago \citep{1988MNRAS.235..715E} and if the improvement
had been due only to computing power then the largest simulation possible
today should have been around $10^9$ particles, whereas the largest simulations
done till date used more than $10^{10}$ particles \citep{2005Natur.435..629S}. 
Evidently, development of new methods and optimisations has also played a
significant role in the evolution of simulation studies
\citep{1985ApJS...57..241E},
\citep{1986Natur.324..446B,1987JCoPh..73..325G,1988ApJS...68..521B,1989ApJS...71..871J,1990JCoPh..87..137H,1990JCoPh..87..148M,1991PASJ...43..621M,1991ApJS...75..231H,1991ApJ...368L..23C,1993PASJ...45..269E,1994CoPhC..78..238T,1995ApJ...453..566B,1995MNRAS.274..287S,1995ApJS...98..355X,1996NewA....1..133D,1997ApJS..111...73K,1998NewA....3..687M,2000ApJS..128..561B,2000NewA....5..163B,2000ApJ...536L..39D,2001MNRAS.325..845K,2001NewA....6...79S,2001ApJ...550L.143K,2002NewA....7..373M,2002JCoPh.179...27D,2002JApA...23..185B,2003NewA....8..665B,2003PASJ...55.1163M,2003ApJS..145....1B,2004astro.ph..5220R,2004NewA....9..111D,2004PASJ...56..521M,2005MNRAS.364.1105S,2005NewA...10..393M,2005PASJ...57..849Y,2004NewA....9..137W,2006CoPhC.174..540T} 
Along the way, code developers have also successfully met the challange posed
by the emergence of distributed parallel programming.

In this paper, we discuss the performance characteristics of an optimisation
for tree codes suggested by Barnes \citep{1990JCoPh..87..161B}.  
We do this in the context of the TreePM method
\citep{2002JApA...23..185B,2003NewA....8..665B} where the tree method is used
for computing the short range force. 
The TreePM method brings in an additional scale into the problem, i.e., the
scale upto which the short range force is computed and this leads to
non-trivial variations in error in force.

The paper is organised as follows: we introduce the TreePM method in \S{2},
and discuss the optimisation scheme in \S{3}.
Performance of the optimisation scheme is discussed in \S{4}, and we discuss
combining this with individual time steps for particles in \S{5}.  
We end with a discussion in \S{6}.

\section{The TreePM Algorithm}

The TreePM algorithm \citep{2002JApA...23..185B,2003NewA....8..665B} is a
hybrid N-Body method which combines the BH-Tree method
\citep{1986Natur.324..446B} with the PM method
\citep{1997Prama..49..161B,2005NewA...10..393M},
\citep{1983MNRAS.204..891K,1983ApJ...270..390M,1985ApJ...299....1B,1985A&A...144..413B,1988csup.book.....H}. 
The TreePM method explicitly breaks the potential into a short-range and a
longe-range component at a scale $r_s$. 
The PM method is used to calculate long-range force and the short-range force
is computed using the BH Tree method.  
Use of the BH Tree for short-range force calculation enhances the force 
resolution as compared to the PM method. 

The gravitational force is divided into a long range and a short range part
using partitioning of unity in the Poisson equation.
\begin{eqnarray}
\phi_k &=& -\frac{4\pi G\rho_k}{k^2} \nonumber\\
&=&  -\frac{4\pi G\rho_k}{k^2} \exp \left(-k^2r_s^2\right) 
-\frac{4\pi G\rho_k}{k^2} 
\left[1- \exp \left(-k^2r_s^2\right) \right] \nonumber\\
&=&  \phi_k^{lr} + \phi_k^{sr} \nonumber \\
\phi_k^{lr} &=& -\frac{4\pi G\rho_k}{k^2} \exp \left(-k^2r_s^2\right) \\
\phi_k^{sr} &=& -\frac{4\pi G\rho_k}{k^2} 
\left[1- \exp \left(-k^2r_s^2\right)\right] 
\end{eqnarray}
Here $\phi_k^{sr}$ and $\phi_k^{lr}$ are the short-range and long-range
potentials in Fourier space. $\rho$ is the density, G is the gravitational 
coupling constant and $r_s$ is the scale at which the splitting of the 
potential is done. The long-range force is solved in 
Fourier space with the PM method and the short-range force is solved 
in real space with the Tree method. The short range force in real space is:
\begin{equation}
\mathbf{f}^{sr}(\br) = -\frac{GM\br}{r^3}
\left[
  \mbox{erfc}\left(\frac{r}{2r_s}\right) + \frac{r}{r_s\sqrt{\pi}}
  \exp\left(-\frac{r^2}{4r^2_s} \right)
\right]
\end{equation}
Here $\mbox{erfc}$ is the complementary error function. 

The short range force is below $1\%$ of the total force at $r \geq 5 r_s$. 
The short range force is therefore computed within a sphere of radius $r_{cut}
\simeq  5 r_s$. 
The short range force is computed using the BH tree method.  
The tree structure is built out of cells and particles.  
Cells may contain smaller cells (subcells) within them.  
Subcells can have even smaller cells within them, or they can contain a
particle. 
In three dimensions, each cubic cell is divided into eight cubic
subcells.  
Cells, as structures, have attributes like total mass, location of centre of
mass and pointers to subcells.  
Particles, on the other hand have the usual attributes: position, velocity and
mass. 

Force on a particle is computed by adding contribution of other
particles or of cells.  
A cell that is sufficiently far away can be considered as a single entity and
we can add the force due to the total mass contained in the cell from its
centre of mass.  
If the cell is not sufficiently far away then we must consider its
constituents, subcells and particles.  
Whether a cell can be accepted as a single entity for force calculation is
decided by the cell acceptance criterion (CAC).  
We compute the ratio of the size of the cell $d$ and the distance $r$ from the
particle in question to its centre of mass and compare it with a threshold
value 
\begin{equation}
\theta = \frac{d}{r} \leq \theta_c  
\end{equation}
The error in force increases with $\theta_c$.  
Poor choice of $\theta_c$ can lead to significant errors
\citep{1994JCoPh.111..136S}. 
Many different approaches have been tried for the CAC in order to minimize
error as well as CPU time usage
\citep{1994JCoPh.111..136S,2001NewA....6...79S}. 
The tree code gains over direct summation as the number of contributions to
the force is much smaller than the number of particles.

The TreePM method is characterised therefore by three parameters, 
$r_s$,$r_{cut}$ and $\theta_c$. 
For a discussion on the optimum choice of these parameters the reader is
referred to \citep{2003NewA....8..665B}.

\section{The scheme of groups}

We first describe an optimization scheme due to \citep{1990JCoPh..87..161B},
given in the paper with a curious title {\sl A modified tree code. Don't
  laugh, it runs}. 
This scheme is easily portable to any N-body algorithm that uses  tree data
structures to compute forces. 
The origin of the optimisation is in the realisation that the tree walk used
for computing forces is computationally the most expensive component of a tree
code. 
The idea is to have a common interaction list for a \emph{group} of particles
that is sufficiently small.
Given that we are working with a tree code, it is natural to identify a cell
in the tree structure as a group. 
One can then add the contribution of particles within the group using direct
pair summation. 
The cell acceptance criterion (CAC) for the tree walk needs to be
modified in order to take the finite size of the group into account. 
In our implementation of the TreePM method, we modified the standard CAC in the
following manner:
\begin{equation}
d \leq \left(r - r_m \right) \theta_c
\end{equation}
where $r_m$ is the distance between the centre of mass of the group, and the
group member that is farthest from the centre of mass. 
This is calculated once before the force calculation and does not add much in
terms of overhead.

The modified CAC can be thought of as the standard CAC with a distance
dependent $\theta_c$, with the value of $\theta_c$ decreasing at small $r$. 
As we require a larger number of operations for smaller $\theta_c$, each tree
walk with the modified CAC is expected to require more CPU time than a tree
walk with the standard CAC.  
However, as we do a tree walk for a group of particles in one go, CPU time is
saved as the time taken for tree walk per particle comes down.

There is an overhead as there is a pair-wise force calculation within
the group.
The cost of this overhead increases as the square of the number of particles
in the group.  
In order to keep the overhead small, one would like the group to be
sufficiently small compared to the size of the N-Body simulation and hence a
maximum size $c_{max}$ and an upper bound on the number of particles in the
group $\npmax$ is used. 
An upper limit on the size of the group is pertinent because of the indirect
effect through the change in the CAC. The effect of the additional 
parameter $\cmax$ with the modified CAC will be seen when we discuss errors 
in section 4. Our implementation of the modified method 
by using a different definition of groups, with the additional 
parameter $\cmax$ and the modified CAC (eq.5) ensure that the short-range 
force is extremely accurate. 
This is different from previous implementations
\citep{1990JCoPh..87..161B,1991PASJ...43..621M,2005PASJ...57..849Y,2004NewA....9..137W}  
where the group scheme was parametrized by just one parameter 
$\npmax$ and the standard CAC (eq.4) used for tree traversal. 
We note in passing that the modified CAC is crucial in order to limit errors.  
Indeed, we find that working with the standard CAC leads to errors in short
range force that are orders of magnitude larger.

\subsection{Estimating Speedup}

We model the modified Tree/TreePM method with the aim of estimating the
speedup that can be achieved.  
If $N$ is the total number of particles, $n_p$ the typical number of
particles in a group and $n_g$ the number of groups then clearly we expect
$n_g \times n_p = N$. 
The total time required for force calculation is a sum of the time taken up
by the tree walk and the time taken up by pair wise calculation within the
group. 
Actual calculation of the force, once the interaction list has been prepared
takes very little time and can be ignored in this estimate, as can the time
taken to construct the tree structure. 
The time taken is:
\begin{equation}
T_g = \alpha n_g \ln{N} + \beta n_g n_p^2 = \alpha
\frac{N}{n_p} \ln{N} + \beta N n_p 
\end{equation}
Here we have assumed that the time taken per tree walk scales as
$\mathcal{O}(\ln{N})$ even with the modified CAC\footnote{This is an
  approximation as we expect the tree walk to depend on $c_{max}$,
  $\npmax$ and $\theta_c$ as well. 
The finite size of groups should lead to deviations from the
$\mathcal{O}(\ln{N})$ variation and the deviation should scale as the ratio of
the volume of the group and the volume of the simulation box. 
As this ratio becomes smaller for large simulation boxes, we feel that the
approximation we have made is justified.}.
The time taken is smallest when
\begin{equation}
n_p = \left(\frac{\alpha \ln{N}}{\beta}\right)^{1/2} ~~~~~~ ; ~~~~~~~~
T_{g_{min}} = 2 \beta N n_p = 2 \alpha \frac{N}{n_p} \ln{N}
\end{equation}
Thus the optimum number of particles in the group scales weakly with the total
number of particles.  
In the optimum situation, we expect the tree walk and the pair wise components
to take the same amount of CPU time.  

For comparison, the time taken for force calculation in the standard TreePM
is: 
\begin{equation}
T = \alpha N \ln{N} 
\end{equation}
and we make the simplifying assumption that $\alpha$ is same for the
two cases.
The expected speed up is then given by:
\begin{equation}
\frac{T}{T_{g_{min}}} = \frac{1}{2} \left(\frac{\alpha
    \ln{N}}{\beta}\right)^{1/2} ~~~~~~~~ .
\end{equation}
The speedup for the optimum configuration scales in the same manner as the
typical number of particles per group.

A more detailed analysis of this type can be found in
\citep{1991PASJ...43..621M}. 

The calculation we have presented above is approximate and ignores several
factors, some of these have already been highlighted above. 
There are other subtleties like the role played by the finite range $r_{cut}$
over which the short range force is calculated. 
The size of a group ($c_{max}$) cannot be varied continuously, and hence $n_p$
is also restricted to a range of values.
Further, the number of operations do not translate directly into CPU time as
some calculations make optimal use of the capabilities of a CPU while others
do not. 
For example, the pair wise calculation is likely to fare better on processors
with a deep pipeline for execution whereas tree walk can not exploit this
feature.  
The finite bandwidth of the CPU-memory connection also has an impact on the
scaling with $N$ for large $N$. 
In the following section, we discuss the implementation of the modified TreePM
method and the timing of the code with different values of parameters. 

\section{A Modified TreePM Algorithm with the Scheme of Groups}

Tests of the TreePM method have shown that $95-98\%$ of the time goes into 
the short-range force calculation. 
Keeping this in mind the scheme of groups was introduced to optimize the
short-range force calculation in terms of speed. 
A welcome feature is more accurate force computation.
Since the optimum set of TreePM parameters have been discussed in
\citep{2003NewA....8..665B}, we now look for the optimum choice of the  
additional parameters, $c_{max}$ and $\npmax$, which describe 
the Modified TreePM algorithm. 
The analysis that follows is divided into two parts. 
First we look at the optimum values of  $c_{max}$ and $\npmax$ which
minimise the time for short range force computation. 
Second, we study errors in total and short range force with this new scheme. 

\subsection{Optimum Parameters of The Modified TreePM Algorithm}

\begin{figure*}
\begin{center}
\begin{tabular}{cc}
\includegraphics[width=2.8truein]{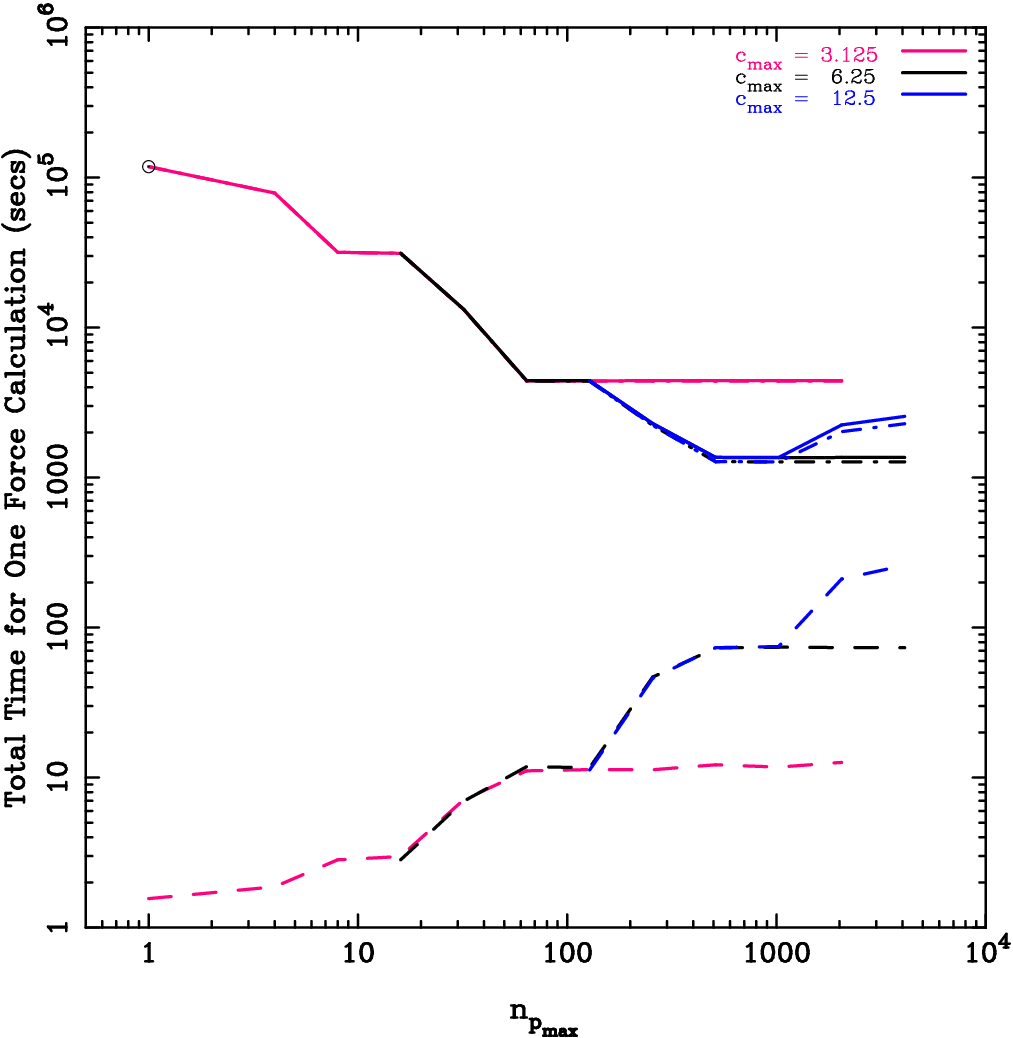} &
\includegraphics[width=2.8truein]{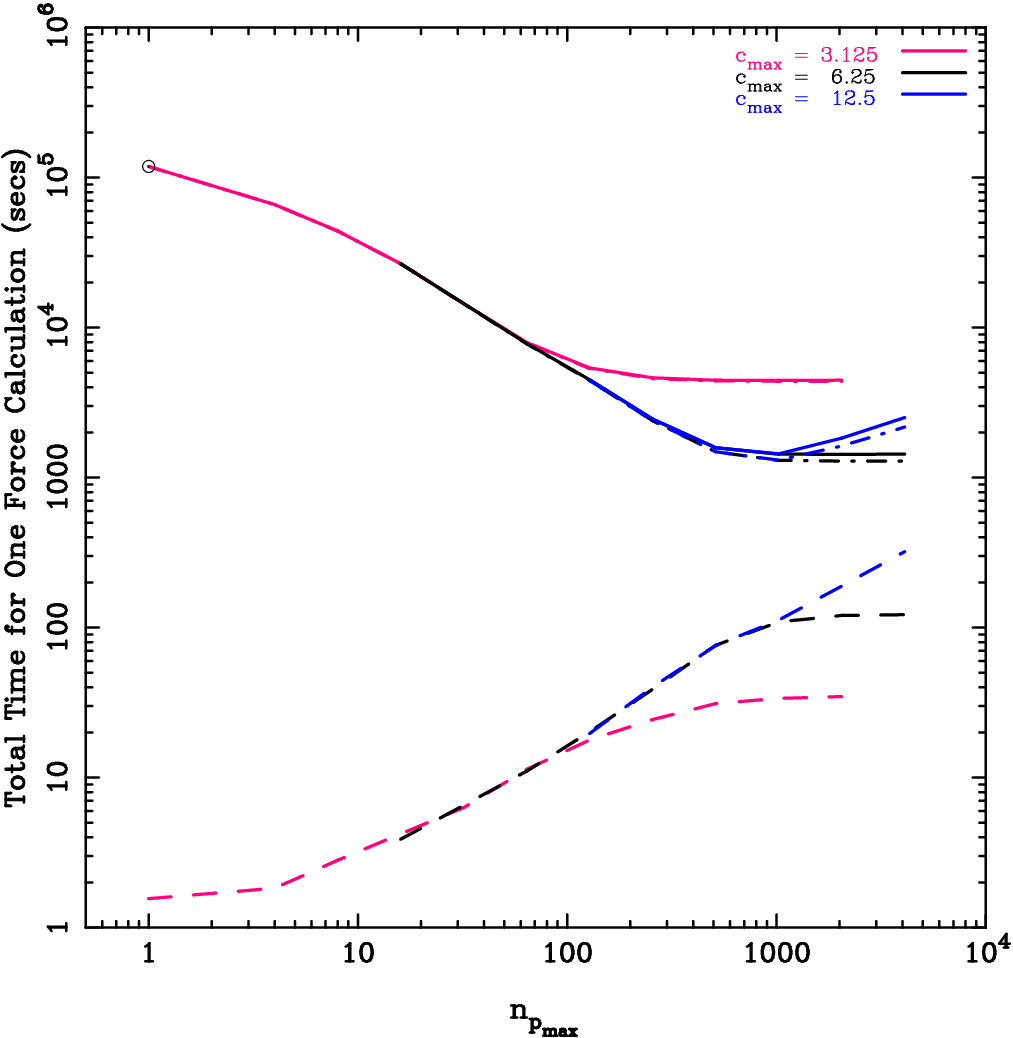}
\end{tabular}
\end{center}
\caption{Time taken for computation of the short term force in the Modified
  TreePM method for an unclustered (left panel) and a clustered (right panel)
  distribution.  Solid lines represent the time taken by a complete
  short-range force calculation.  Dashed lines are the contribution to the
  force due to pairs within a group, the intra-group contribution.  Dot-Dashed
  lines are the contributions to force due to tree walk. Purple, black
  and blue lines are for $\cmax = 3.125,6.25,12.5$ respectively} 
\end{figure*}

We choose $r_s = 1$, $r_{cut} = 5.2r_s$ and $\theta_c = 0.5$ for the
discussion that follows. 
With this choice the error in force for  $99\%$ of the particles is less than
a few percent \citep{2003NewA....8..665B}. 
We present analysis of performance of the modified TreePM for two 
different particle distributions taken from an $N-$body simulation, with 
$N = N_{box}^3 = 200^3$.
\begin{itemize}
\item
An unclustered distribution that corresponds to the initial conditions of an
$N-$body simulation. 
\item
A clustered distribution taken again from the same $N-$body simulation.  
The scale of non-linearity for the clustered distribution is $8$ grid lengths.
\end{itemize}
We have verified that the nature of results does not change significantly for
simulations with the number of particles ranging from $32^3$ to $256^3$.

In figure 1 we show the time taken for computing the the short-range force
(solid line) and determine the values of ($c_{max},\npmax$) for which
this timing is a minimum. 
Two leading contributions to the calculation of short range force are shown
separately: 
\begin{itemize}
\item
Intra-group particle-particle contribution (dashed line).
\item
Time take for tree-walk and the related force calculation (dot-dashed line). 
\end{itemize}

Given that a group is a cell with maximum width  
\begin{equation}
c_{max} = \frac{N_{box}}{2^m}, \,\,\,\,\,\,\mbox{($m$ is an integer)},
\end{equation}
$c_{max}$ can take therefore only discrete values. 
We choose to restrict upto $c_{max} \sim 2r_{cut}$ as for larger cells, the
dominant contribution to force on a given particle arises from the intra-group
particle-particle interaction and the time taken for this is a sensitive
function of the amplitude of clustering. 

The time taken for computing the short range force in both, unclustered (left
panel) as well as clustered (right panel), distributions is qualitatively
described by our model (see eqn.(6)). 
The pairwise force increases linearly with $\npmax$, $\npmax$ is the maximum
number of particles in a group and scales as $n_p$ where $n_p$ is the average
number of particles in a group. 
The time taken for tree-walk decreases as  $\npmax^{-0.65}$, reaches a minimum
and then increases with $\npmax$ (blue line) for the largest $\cmax$ used
here.
For other values of $\cmax$ we see the timing levelling off near the minimum. 
The scaling as $\npmax^{-0.65}$ is different from $1/n_p$ we used in the
analytical model and the reason for this is likely to be in the approximations
we used. 
We find that the scaling approaches $1/\npmax$ as we consider simulations with
a larger number of particles. 
One crucial reason for different scaling is the modified CAC we use here.  
This effectively leads to a smaller $\theta_c$ for cells closer to the group
and the number of such cells increases with $\cmax$.

In both cases the total time is still dominated by the tree-walk. 
The plateaus in the plots often indicate that the number of particles in a
group of maximum size $\cmax$ have saturated. 
At initial times where the fluctuations are small there is also a lower bound
on the number of particles contained in a group. 
In the clustered distribution there is no such lower bound but an upper bound,
larger than the corresponding upper bound in case of the unclustered
distribution, exists and is dictated by the amplitude of clustering in the
distribution of particles.

From figure~1 we see that the optimum values of ($\cmax,\npmax$) =
($12.5$, $1024$) $\&$ ($6.25$, $\geq 1024$) given by the minima of the solid
blue line and the plateau of the solid black line respectively.
In the latter case the time taken does not change for $\npmax \geq 1024$ and
we consider this to be a useful feature that makes the $\cmax = 6.25$ a better
choice as fine tuning of $\npmax$ is not required.
For the optimum ($\cmax$, $\npmax$) one can see that force computation takes
the same time for the clustered and the unclustered distributions. 
Table~1 lists optimum values for ($\cmax$, $\npmax$) for N-Body simulations
with different numbers of particles. 
These numbers indicate that a good choice for $\cmax$ is one 
which is closest to $r_{cut}$, i.e.  $\cmax \sim r_{cut}$. 
The parameter $\npmax$ can be taken to be $10^3 \leq \npmax$ as we find little
variation beyond this point.

\begin{table}
\caption{The following table lists the optimum values of ($\cmax,\npmax$) for 
simulations of various sizes but having the same TreePM parameters: $r_s = 1$,
$r_{cut} = 5.2r_s$ $\theta_c = 0.5$.} 
\vspace{0.3cm}
\begin{center} 
\begin{tabular}{||c|l|l||} 
\hline\hline
$N_{box} = N^{1/3}$ 
& $\cmax^{opt}$  
& $\npmax^{opt}$
\\ \hline\hline
64 & 4.0 & $\geq$1024 
\\ \hline
128 & 4.0 & $\geq$1024 
\\ \hline
160 & 5.0 & $\geq$1024 
\\ \hline
200 & 6.25 & $\geq$1000 
\\ \hline
\hline
\end{tabular}
\end{center}
\vspace{0.3cm}
\end{table}

One can get an estimate of the overheads for the group scheme by looking at
the limit of $\npmax \rightarrow 1$. 
Here we compare the performance of the TreePM with the modified code by
plotting the time taken by the former as a large dot on the same panel where
the time taken for the modified code is shown in the form of curves. 
The difference between these timings is around  $0.1\%$.  

The speedup for the optimal configuration of the modified TreePM, as compared
with the base TreePM code is $\sim 83$.
This is a huge gain and has to do with better utilisation of the CPU cache. 
The speedup is less impressive for smaller simulations, and is larger for
bigger simulations. 
This is shown in figure~2 where we plot the time taken for force calculation
per particle per step as a function of the total number of particles in the
simulation. 
This is shown for the TreePM as well as the modified TreePM codes.
Performance on two different types of processors is shown here to demonstrate
that the optimisation works equally well on these.
One can see that the TreePM code becomes (CPU-Memory) bandwidth limited for
simulations with more than $64^3$ particles and the time taken increases more
rapidly than $\mathcal{O}(\ln{N})$. 
This does not happen in case of the modified TreePM where the scaling is
$\mathcal{O}(\ln{N})$ throughout.
It is this difference that leads to impressive speedup for large simulations. 
For simulations with up to $64^3$ particles we get a speedup by a factor of
four. 


\begin{figure}
\begin{center}
\includegraphics[width=3.5truein]{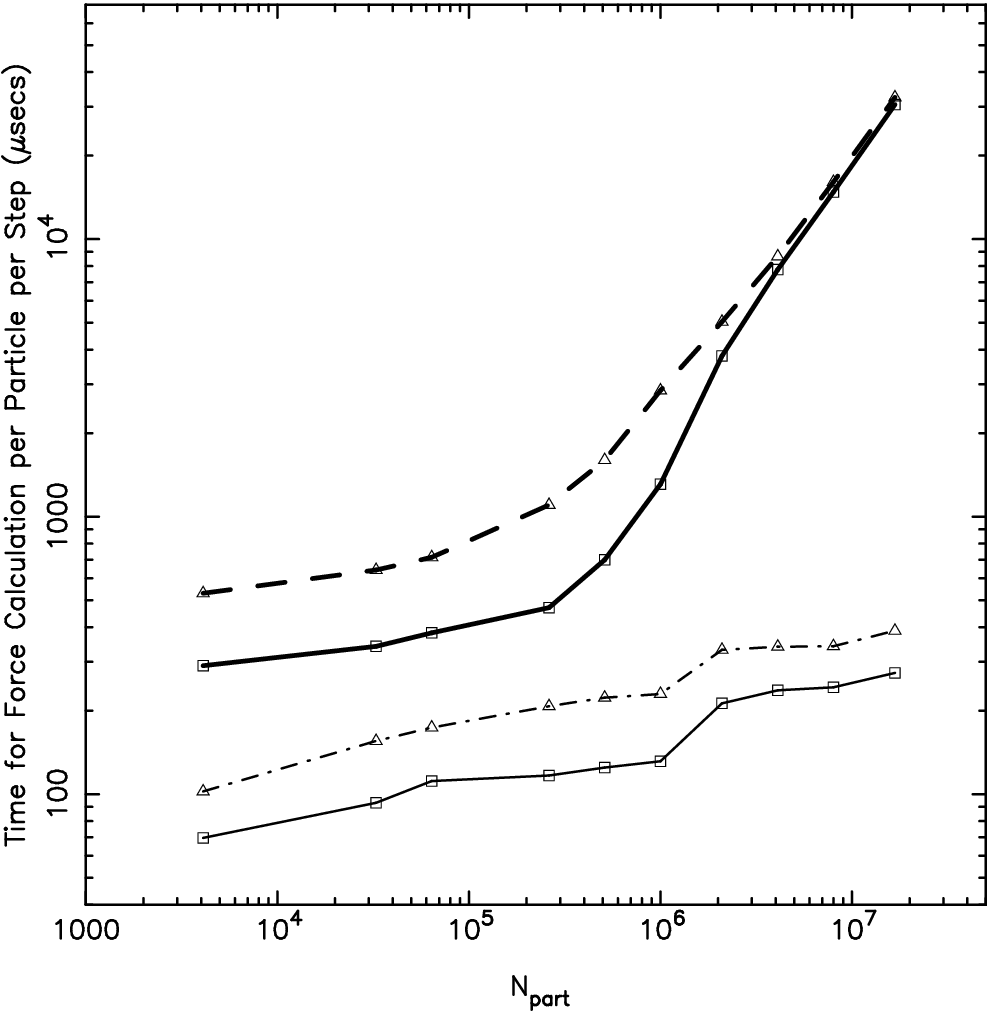}
\end{center}
\caption{Time taken for short-range force calculation per particle per step
  for $N=32^3$ to $N=256^3$ for the TreePM (thick line) and the Modified
  TreePM (thin line).  The solid line shows the performance of the codes on a
  single core if Intel 5160 (3.0 GHz) processor and the dashed line shows the
  performance on a single core of the AMD Barcelona (2.1 GHz) processor.}
\end{figure}

\subsection{Errors in the Modified TreePM Force}

We now study errors in force for the modified TreePM force. 
Errors are calculated with respect to a reference force computed 
with very conservative values of TreePM parameters: 
$\theta_c = 0.01$, $r_s = 4.0$, $r_{cut} = 5.2r_s$. 
With these values the reference force is accurate to $0.1\%$
\citep{2003NewA....8..665B}. 
\begin{equation}
\epsilon = \frac{|\mathbf{f_{ref}} - \mathbf{f}|}{|\mathbf{f_{ref}}|}
\end{equation}
Here $\epsilon$, $\mathbf{f_{ref}}$, $\mathbf{f}$ are the relative error, 
reference force and the typical force in a simulation.  
We calculate errors for two distributions of particles:
\begin{itemize}
\item
A uniform (unclustered) distribution.
\item
A clustered distribution taken from an $N-$body simulation. 
\end{itemize}
Both distributions have $N_{box}^3=N=128^3$ particles.
The exercise we follow is similar to \citep{2003NewA....8..665B} 
but now we wish to highlight the effect of groups on errors in force. 

\begin{figure*}
\begin{center}
\begin{tabular}{cc}
\includegraphics[width=2.8truein]{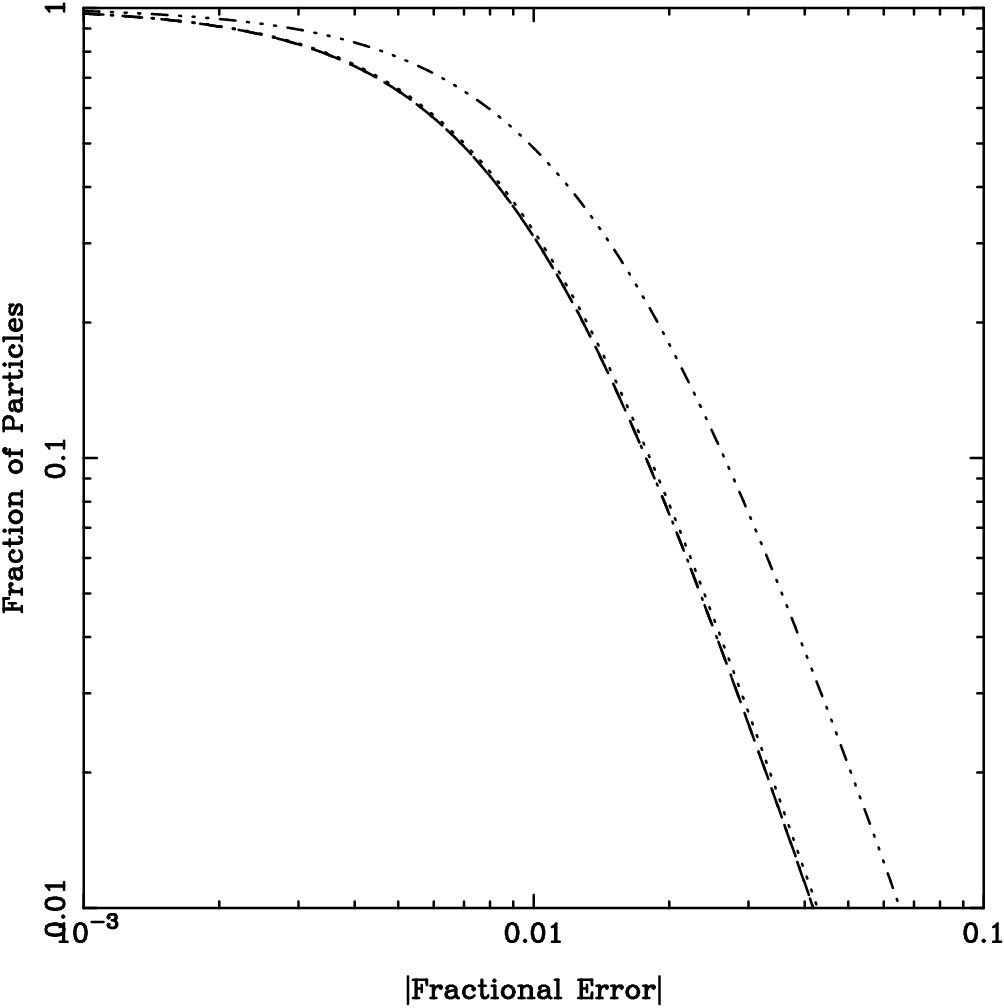} &
\includegraphics[width=2.8truein]{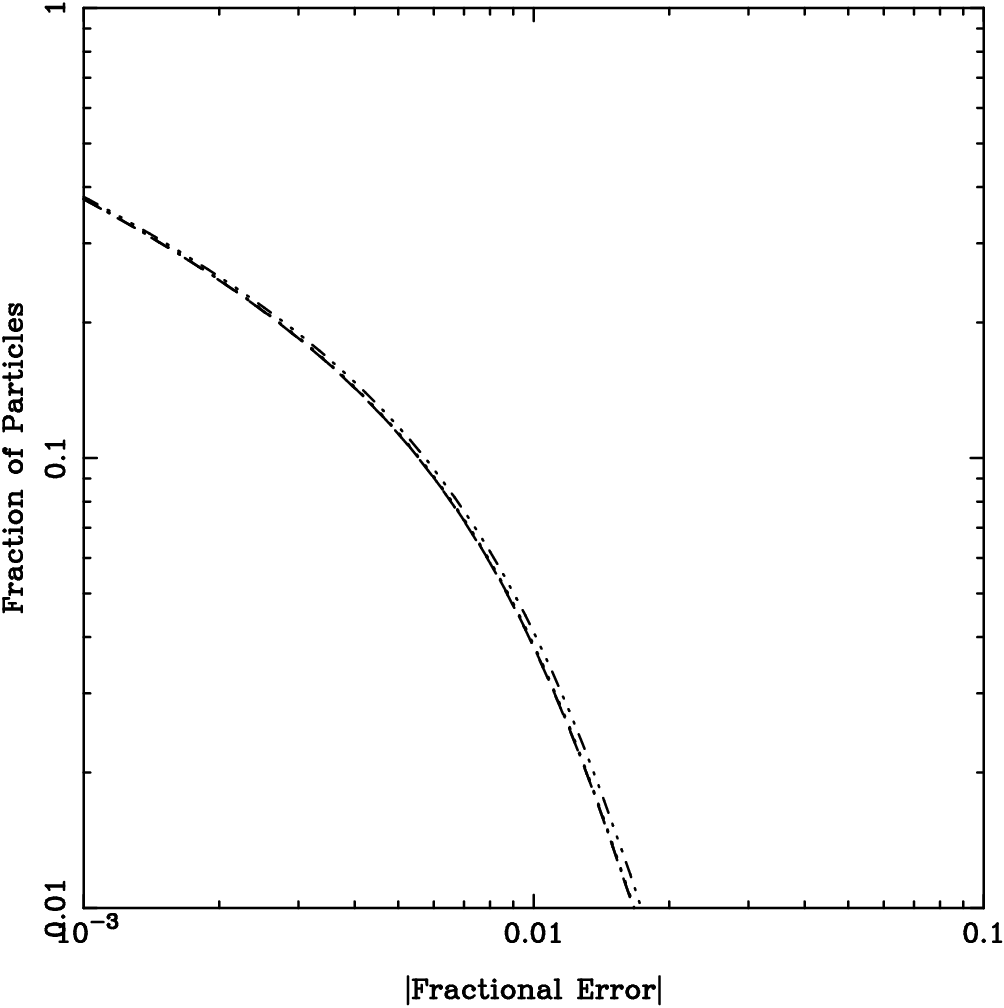} \\
\includegraphics[width=2.8truein]{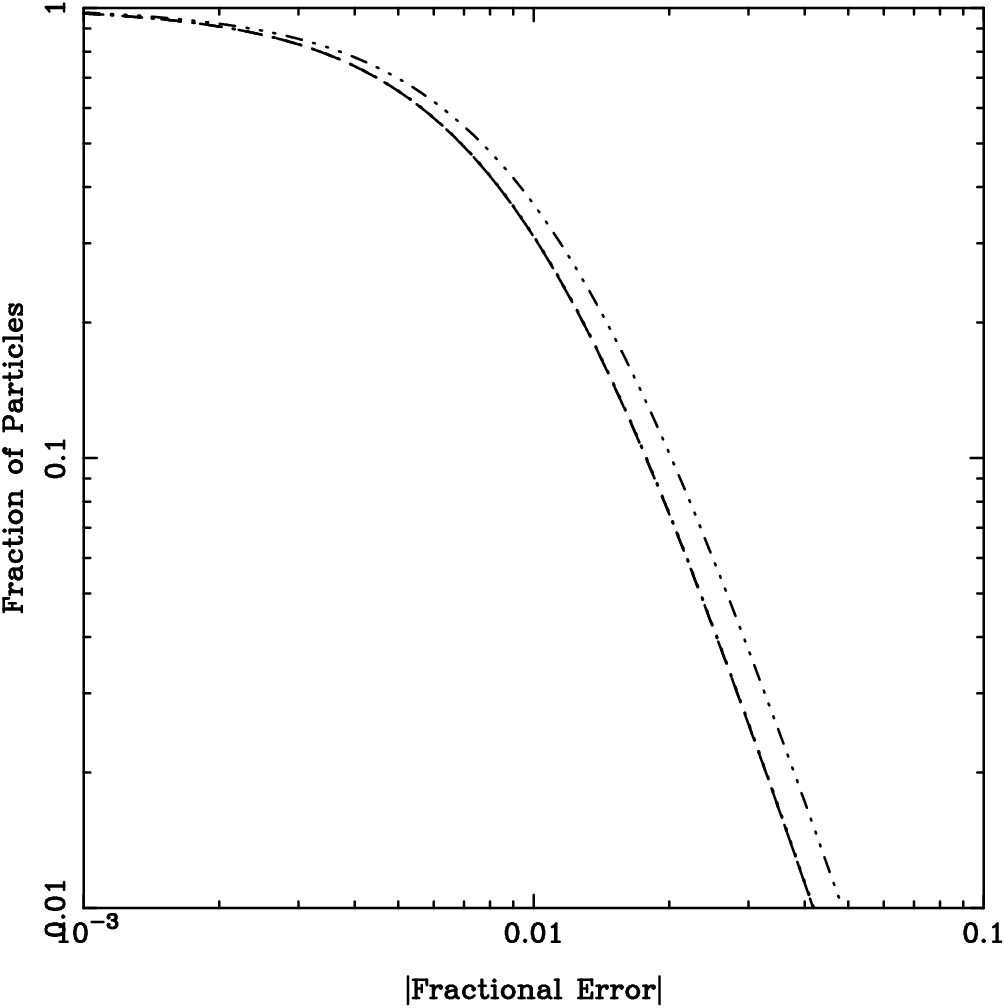} &
\includegraphics[width=2.8truein]{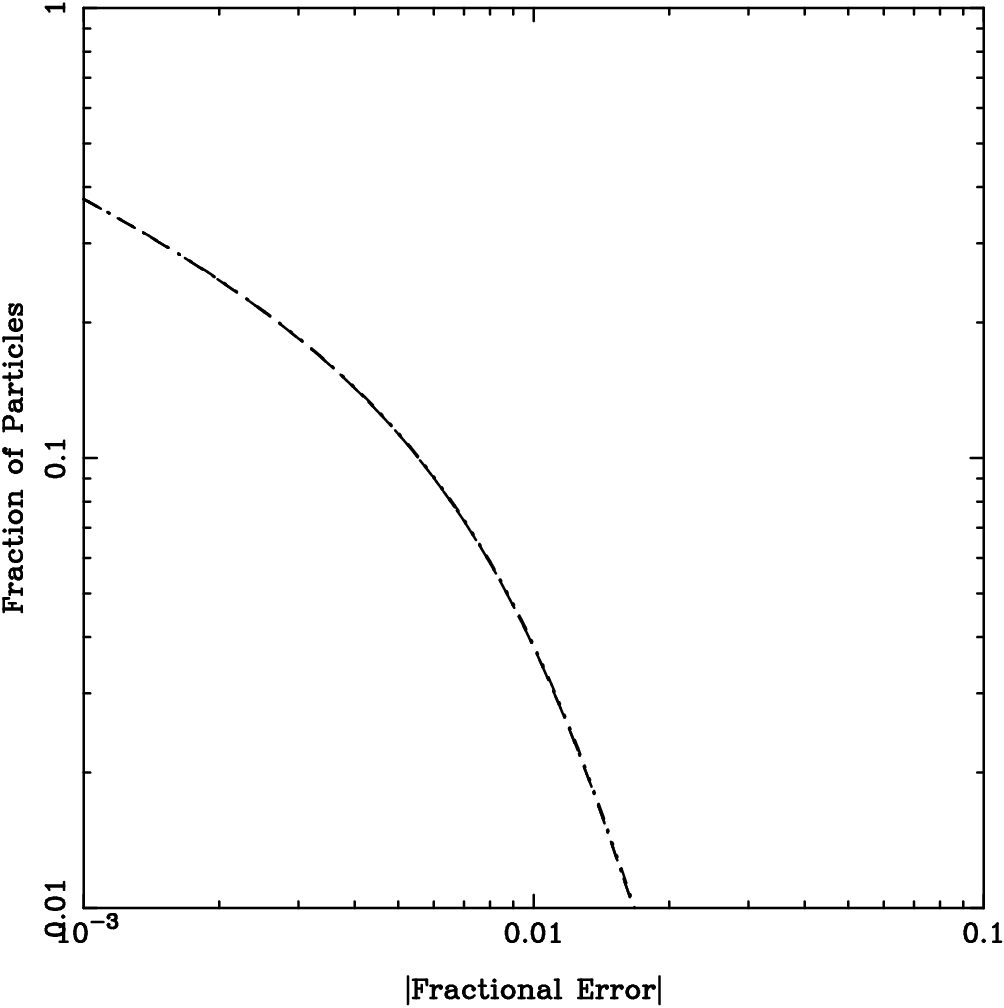}
\end{tabular}
\end{center}
\caption{Distribution of errors in total force for different values 
of $\theta_c$ with $\cmax=2.0$ for unclustered (top left panel) and 
clustered (top right panel) distributions. 
Dashed, dot-dash-dot-dash, dotted, dash-dot-dot-dot 
lines are for $\theta_c=0.1$, $0.2$, $0.3$ and
$0.5$ respectively.  
We used $r_s = 1.0, r_{cut} = 5.2r_s$ for these plots.  
The corresponding plots for $\cmax=4.0$ are shown in lower left and lower
right panels.} 
\end{figure*}

Figure 3 shows the distribution of errors for different values 
of $\theta_c$. 
The results are shown for both the distributions being studied here: the
unclustered distribution (left panels) and the clustered distribution (right
panels). 
The top row is with $\cmax = 2.0$ and the lower row is for $\cmax = 4.0$. 
We used $r_s = 1.0$ and $r_{cut} = 5.2r_s$ for this figure. 
In the case of the unclustered distribution error decreases with $\theta_c$
but saturates at $\theta_c = 0.3$ and does not decrease as $\theta_c$ is
decreased further. 
The situation is different for the clustered distribution where the errors are
not sensitive to $\theta_c$.  
This suggests that the errors are dominated by the long-range force.
The unclustered distribution has larger errors than the 
clustered distribution. 
This is because the net force on each particle in the unclustered distribution
is small, whereas force due to a cell with many particles is large and many
such large contributions have to cancel out to give a small net force. 
Numerical errors in adding and subtracting these large numbers seem to
systematically give a net large error. 
Larger cells contribute for larger $\theta_c$ hence the variation with
$\theta_c$ is more dramatic in the unclustered case. 
This effect is apparent in the discussion of the short-range force.  
With $\theta_c = 0.3$, $1\%$ of particles have errors in total force greater
than $4\%$ in the unclustered case and $1.6\%$ in clustered case. 

The effect of the modified CAC (eqn. (5)) is seen by comparing the plots of
figure~3 for the unclustered distribution. 
The modified CAC is more stringent for larger value of $\cmax$ and this is
clearly seen in the error for $\theta_c = 0.5$. 
There is a lack of variation in errors with $\theta_c$ for $\theta_c <
0.5$ indicating that at this stage the dominant contribution to errors is from
the long range force calculation.
The short-range force is more accurate with a larger $\cmax$ due to two
reasons:
\begin{itemize}
\item
The modified CAC has an $r$ dependent opening angle threshold and requires a
smaller $\theta_c$ at small distances.  
This is likely to reduce errors.
\item
The number of particles in a group is larger for larger $\cmax$.  
As the contribution of force from these particles is computed by a direct
summation over pairs, the errors are negligible.
\end{itemize}
One may raise the concern that the errors in the present approach are likely
to depend on location of a particle within the group.
We have checked for anisotropies in error in force calculation in groups that
may result and we do not find any noteworthy anisotropic component.

\begin{figure*}
\begin{center}
\begin{tabular}{cc}
\includegraphics[width=2.8truein]{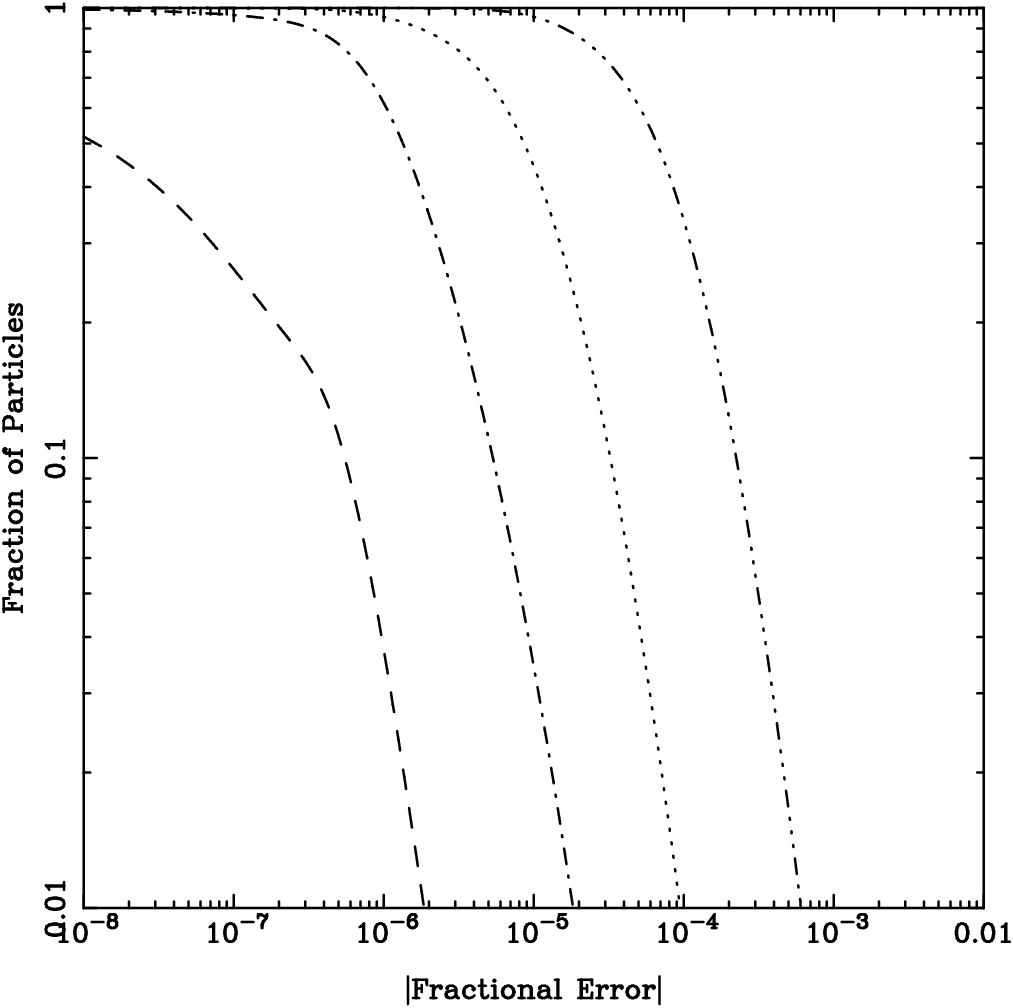} &
\includegraphics[width=2.8truein]{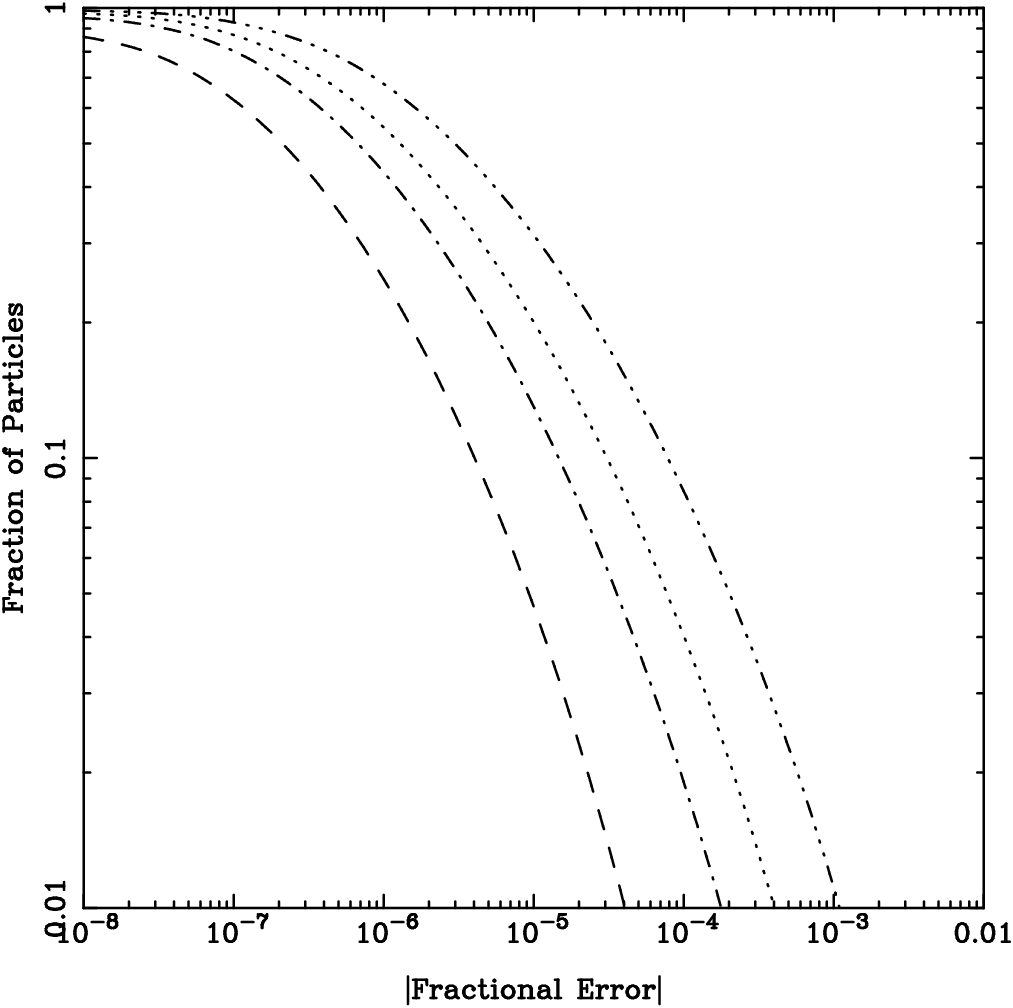}
\end{tabular}
\end{center}
\caption{Distribution of errors in short range force for different values 
of $\theta_c$ for unclustered (left panel) and clustered (right panel)
distributions.
Dashed, dot-dash-dot-dash, dotted, dash-dot-dot-dot 
lines are for $\theta_c=0.1$, $0.2$, $0.3$ and
$0.5$ respectively.
$\cmax = 4.0$, $r_s = 1.0$, $r_{cut} = 5.2r_s$ was used for these plots.}
\end{figure*}

Next we look at the errors in short-range force for the same distributions 
(unclustered and clustered) of particles for various values of $\theta_c$. 
The reference short-range force was computed with $\theta_c=0.01$, $r_s=1.0$,
$r_{cut}=5.2r_s$ and $\cmax=4.0$.  
We only varied $\theta_c$ and continued to use $r_s=1.0$, $r_{cut}=5.2r_s$,
$\cmax=4.0$ for computing the short-range force and then the errors. 
For the purpose of computing errors in the short range force, we cannot vary
$r_s$ between the reference and the test model.

The effect of decreasing $\theta_c$ is more dramatic on errors in the
short-range force. 
For the unclustered case the errors for $1\%$ of the particles decreases by
nearly $2.5$ decades to $2\times10^{-4}\%$ for $\theta_c=0.1$. 
In the clustered case  the errors for $1\%$ of the particles decrease by
nearly $1.5$ decades to $3.4\times10^{-2}\%$ for $\theta_c=0.1$. 
One can obtain very high accuracy in short-range force by taking $\theta_c =
0.2$. 
As the short range force is the dominant one at small scales, the TreePM code
can be used to follow the local dynamics fairly well by using a smaller
$\theta_c$. 
The impact of a small $\theta_c$ on CPU time remains to be seen though.

\begin{figure*}
\begin{center}
\begin{tabular}{cc}
\includegraphics[width=2.8truein]{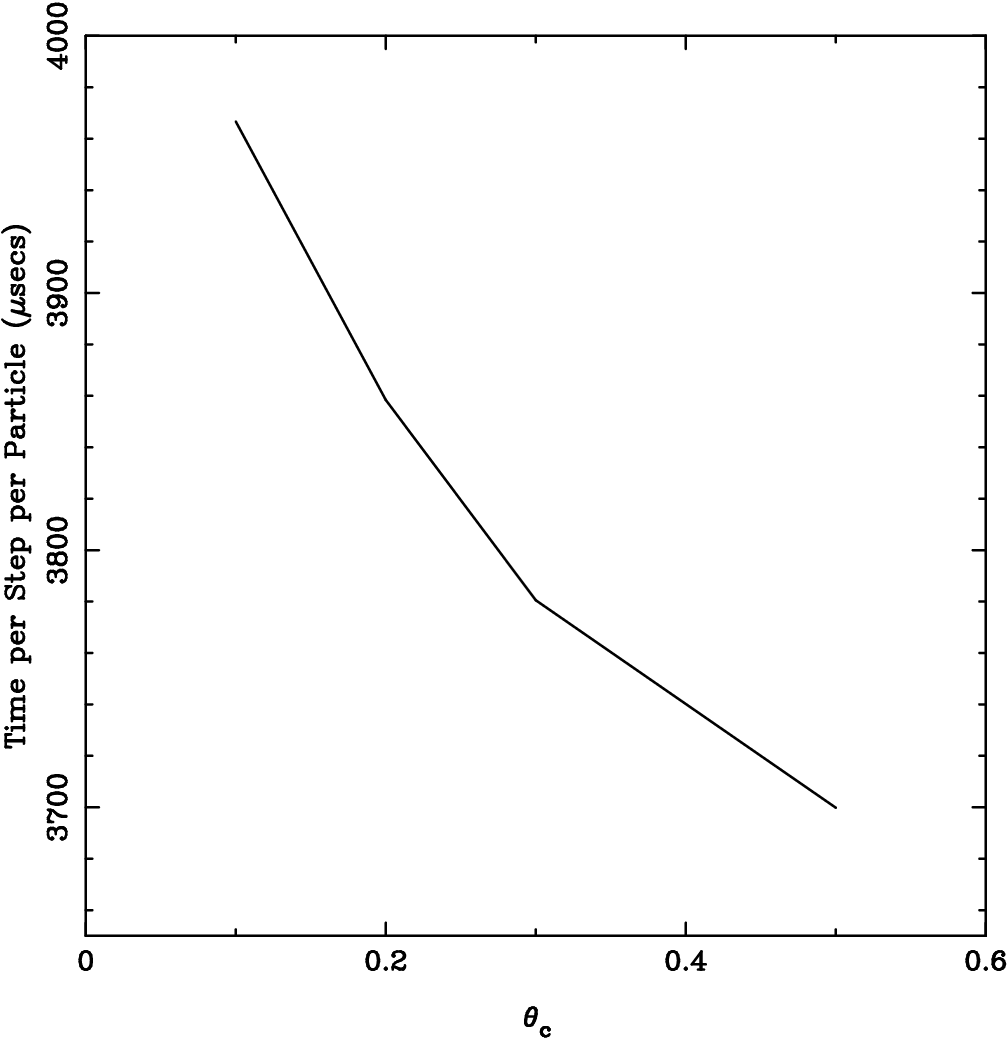} &
\includegraphics[width=2.8truein]{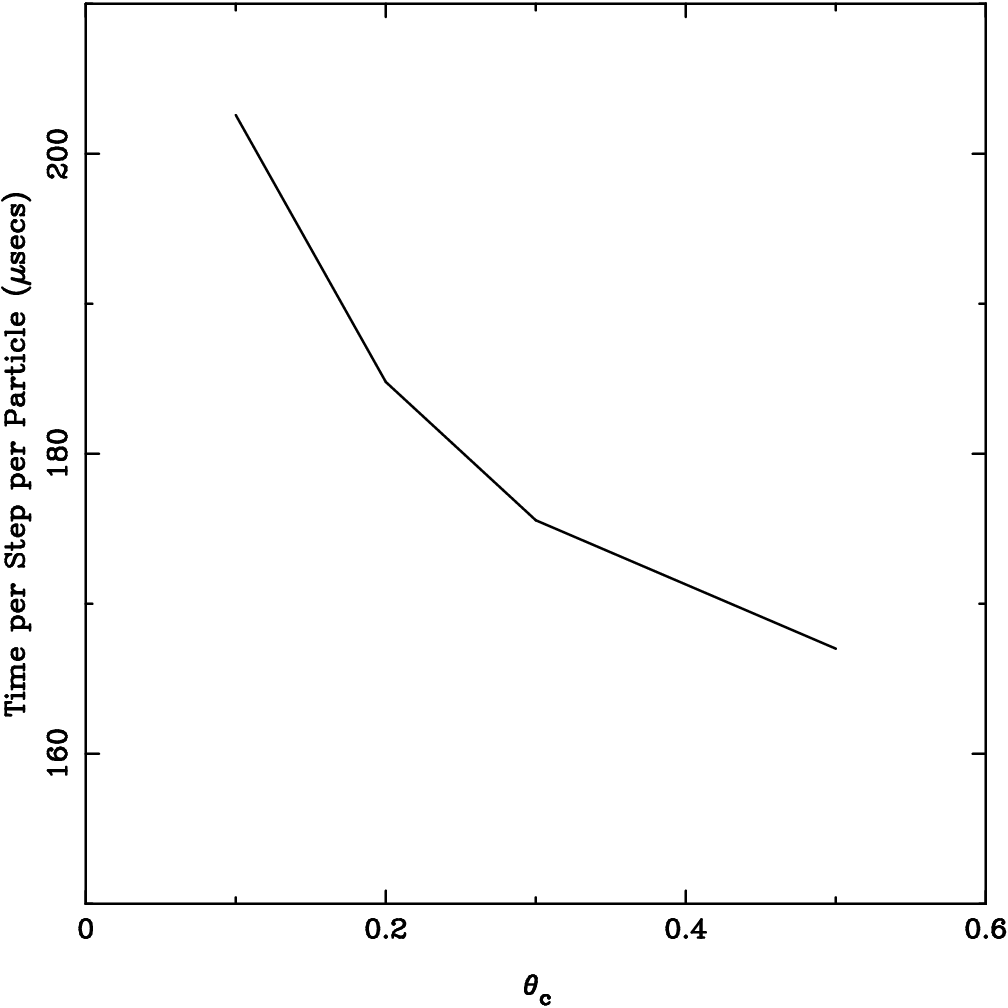}
\end{tabular}
\end{center}
\caption{Scaling of the time taken for short-range force calculation with
  $\theta_c$ for the TreePM (left panel) and the modified TreePM (right
  panel).} 
\end{figure*}

In figure 5 we look at how the CPU time for force calculation scales with
$\theta_c$ for the TreePM (left panel) and the modified TreePM (right panel). 
We compute the time taken for short-range force calculation per particle per
timestep.  
We have seen in figure~1 and the corresponding discussion that clustering does
not seriously affect the performance of the TreePM code. 
We therefore do not repeat the exercice for distributions with different
levels of clustering.  
We performed the short-range force timing on a clustered distribution taken
from an $N-$body simulation with $N_{box}^3=N=128^3$.   
We used $r_s=1.0$ and $r_{cut}=4.0$ for the TreePM and the modified TreePM. 
In addition $\cmax=4.0$, $\npmax=1024$ were used for timing the modified
TreePM. 
When $\theta_c$ is decreased from $0.5$ to $0.2$ the time for force
computation per particle increases by $7.2\%$ for the TreePM and $21\%$ for
the modified TreePM.  
The speedup of the modified TreePM over the TreePM when $\theta_c$ is reduced
from $0.5$ to $0.2$ decreases from $22.2$ to $19.6$, respectively. 
A nice feature of TreePM codes is that unlike tree codes, the CPU time taken
by TreePM codes is less sensitive to $\theta_c$\footnote{For example, the
  variation in CPU time for a tree code increases by about $500\%$ for the
  same change in $\theta_c$ for a simulation with $N \approx 10^4$, and the
  increase in CPU time is larger for simulations with a larger number of
  particles \citep{1987ApJS...64..715H}.}. 
Thus one can obtain much higher accuracy for the short range force with a
TreePM code for a considerably smaller cost in terms of the CPU time.

\section{A Hierarchy of Timesteps}

Due to the existence of a large range of dynamical time scales in a simulation
of large scale structures, computing forces for slowly moving particles at
every timestep is not required. 
It is better to integrate the orbits of rapidly moving particles with a
smaller timestep than those that move relatively slowly, this reduces the
number of force calculations that are required.
As force calculation is the most time consuming component of an N-Body code,
this results in a significant reduction of the CPU time required. 
We have implemented a hierarchical time integrator similar to that used in
GADGET-2 \citep{2005MNRAS.364.1105S}, in which particle trajectories are
integrated with individual timesteps and synchronised with the largest
timestep. 
As we allow the block time step\footnote{Same as the largest time step.} to
vary with time, we work with the so called KDK approach (Kick-Drift-Kick) in
which velocties are updated in two half steps whereas position is updated in a
full step.
It can be shown that with a variable time step, KDK performs better than DKD
(Drift-Kick-Drift) (see the GADGET-2 paper \citep{2005MNRAS.364.1105S} for
details.). 
In our implementation of the hierarchy of time steps, the smaller time steps
differ by an integer power ($n$) of $2$ from the largest, block time step. 
An array is then used to store the value $n$ which determines the timestep of
the particle. 
The code drifts all the particles with the smallest timestep to the next time,
where a force computation is done for particles that require an updation of
velocity (Kick). 
We have tested the robustness of the hierarchical KDK integrator by
succesfully integrating the $3-$body problem discussed by Szebehely \&
Peters, \citep{1967AJ.....72.1187S}. 

\begin{table*}
\caption{This table lists the time taken for a complete simulation run 
for the unoptimised TreePM, TreePM with hierarcical time steps, 
TreePM with the group scheme, and finally the TreePM with the
group scheme as well as the hierarchical time steps and their speedup
with respect to the base TreePM.}
\vspace{0.3cm}
\begin{center}
\begin{tabular}{||c|c|c|c|c||} 
\hline\hline
Run & Groups & Individual Timesteps & Time (secs)& Speedup w.r.t Run 1
\\ \hline\hline
1 & No  & No  &  401983 & 1.0
\\ \hline
2 & No  & Yes &  145240 & 2.77
\\ \hline
3 & Yes & No  &  67639  & 5.94
\\ \hline
4 & Yes & Yes &  31612  & 12.72
\\ \hline
\hline
\end{tabular}
\end{center}
\vspace{0.3cm}
\end{table*}

Solving the equation of motion with a hierarchy of time steps can be combined
with the group scheme. 
Since tree construction takes a small fraction of the total time, a new tree
can be constructed whenever particles require an updation of velocity. 
The groups that contain such particles can then be identified and particles
within each group can be reordered into two disjoint sets: ones that need an
updation of velocity and others that don't. 
Force is computed only for particles in the first set. 
Since each group represents a very small fraction of the total number of
particles, the overhead of reordering particles is negligible. 

Table~2 lists the time taken for a complete simulation run 
for the unoptimised TreePM, TreePM with hierarcical time steps, 
TreePM with the group scheme, and finally the TreePM with the
group scheme as well as the hierarchical time steps and their speedup
with respect to the base TreePM. 
The model used for this comparison is a power law model with $n=-1.0$,
$N_{box}^3=N=64^3$. 
We used $r_s=1.0$, $r_{cut}=5.2r_s$, $\theta_c=0.5$ and $\epsilon=0.2$ in all
the runs. 
Here $\epsilon$ is the softening length.  
We used $\cmax=4.0$ and $\npmax=1024$ for the modified TreePM.

We note that the hierarchical integrator gives a speedup of better than a
factor $2$, irrespective of whether the scheme of groups is used or not. 
The speedup is larger if the softening $\epsilon$ is smaller, as the number of
levels in the hierarchy increases with decreasing $\epsilon$. 
The scheme of groups on the other hand gives a speedup of $4$ or better for
small simulations, and a much larger speedup for bigger simulations.
This speedup has little dependence on the TreePM parameters, i.e. $\theta_c$,
$r_s$, $r_{cut}$. 
The combination of two optimisations gives us a speedup of $10$ or more even
for small simulations. 

\section{Discussion}

The scheme of groups when combined with a hierarchical integrator for the
equation of motion guarantees a speedup of better than $10$ for any $N-$body
code which uses tree structures for computing forces. 
From the algorithmic point of view one does not expect a much larger speedup
for larger $N$. 
However, as seen in Figure~2, the scheme also allows us to make better use of
the cache on CPUs and the effective speedup can be even more impressive.  
We have demonstrated that memory overhead is negligible, and as was observed in
\citep{1990JCoPh..87..161B} this optimisation just takes around $200$ extra
lines of code. 
A welcome feature is more accurate force computation than the code without
this modification. 
This modification in principle introduces two additional parameters ($\cmax$,
$\npmax$), but these are not independent and we have found that $\cmax \sim
r_{cut}$ and $\npmax \geq 10^3$ are good choices across a range of simulation
sizes. 

Our analysis of the optimisation has been restricted to fixed resolution
simulations. 
In case of zoom-in simulations the range of time scales is much larger and a
more complex approach for combining the group scheme with the hierarchy of
time steps may be required.  
Relative efficacy of the two optimisations may be very different in such a
case when compared with the example studied in the previous section.

In summary, we would like to point out that the scheme of groups leads
to a significant optimisation of the TreePM method.
The amount of CPU time saved is significant even for small
simulations, but the cache optimisation aspect leads to even more
significant gains for large simulations. 
We have shown in this paper, that it is possible to incorporate the
scheme in a simple manner in any tree based code. 
The overall gain is very impressive as we are able to combine this
with the use of a hierarchy of time steps. 
The possibility of combining the two optimisations has been explored
in this work for the first time.


\begin{acknowledgements}
Numerical experiments for this study were carried out at cluster computing
facility in the Harish-Chandra Research Institute
(http://cluster.hri.res.in).
This research has made use of NASA's Astrophysics Data System.
The authors would like to thank Hugh Couchman, Jun Makino and Volker Springel
for useful comments. 
\end{acknowledgements}



\end{document}